\documentclass[prl,twocolumn,superscriptaddress]{revtex4}
\usepackage{graphicx}
\usepackage{amsmath}
\usepackage{amsfonts}
\usepackage{amssymb,mathrsfs}

\newcommand{\bs}[1]{\boldsymbol{#1}}

\begin{document}

\title{Giant nonlocality in nearly compensated 2D semimetals}

\author{S. Danz}
\affiliation{\mbox{Institut f\"ur Theorie der kondensierten Materie, Karlsruhe Institute of
Technology, 76128 Karlsruhe, Germany}}
\author{M. Titov}
\affiliation{Radboud University Nijmegen, Institute for Molecules and Materials, NL-6525 AJ
Nijmegen, The Netherlands}
\affiliation{ITMO University, 197101 St. Petersburg, Russia}
\author{B.N. Narozhny}
\affiliation{\mbox{Institut f\"ur Theorie der kondensierten Materie, Karlsruhe Institute of
Technology, 76128 Karlsruhe, Germany}}
\affiliation{National Research Nuclear University MEPhI (Moscow Engineering Physics Institute),
  115409 Moscow, Russia}

\date{\today}

\begin{abstract}
  In compensated two-component systems in confined, two-dimensional
  geometries, nonlocal response may appear due to external magnetic
  field. Within a phenomenological two-fluid framework, we demonstrate
  the evolution of charge flow profiles and the emergence of a giant
  nonlocal pattern dominating charge transport in magnetic
  field. Applying our approach to the specific case of intrinsic
  graphene, we suggest a simple physical explanation for the
  experimental observation of giant nonlocality. Our results provide
  an intuitive way to predict the outcome of future experiments
  exploring the rich physics of many-body electron systems in confined
  geometries as well as to design possible applications.
\end{abstract}

\maketitle

The trend towards miniaturization of electronic devices requires a
deeper understanding of the electron flow in confined geometries. In
contrast to the electric current in household wiring, charge flow in
small chips with multiple leads may exhibit complex spatial
distribution patterns depending on the external bias, electrostatic
environment, chip geometry, and magnetic field. Traditionally, such
patterns were detected using nonlocal transport measurements
\cite{skocpol,vwees1,geimnl1,roukes,vonklit,nlgold,geimnl2}, i.e. by
measuring voltage drops between various leads other than the source
and drain. Devised to study ballistic propagation of charge carriers
in mesoscopic systems, these techniques were recently applied to
investigate possible hydrodynamic behavior in ultra-pure conductors
\cite{geim1,geim3,geim4,rev,luc}, where the unusual behavior of the
nonlocal resistance is often associated with viscosity of the
electronic system \cite{pol15,fl0,fl1,pol17,sven1}.

Nonlocal resistance measurements have also been used to study edge
states accompanying the quantum Hall effect
\cite{mceu,goldman,roth,nlr,caza,koma}. While the exact nature of the
edge states has been a subject of an intense debate, the nonlocal
resistance, $R_{NL}$, appears to be an intuitively clear consequence
of the fact that the electric current flows along the sample edges and
not through the bulk. Such a current would not be subject to
exponential decay \cite{pauw} exhibited by the bulk charge propagation
leading to a much stronger nonlocal resistance.

In recent years the focus of the experimental work on electronic
transport has been gradually shifting towards measurements at nearly
room temperatures \cite{nlgold,nlr,geim1,geim3,geim4}. A particularly
detailed analysis of the nonlocal resistance in a wide range of
temperatures, carrier densities, and magnetic fields was performed on
graphene samples \cite{nlr}. Remarkably, the nonlocal resistance
measured at charge neutrality remained strong well beyond the quantum
Hall regime, with the peak value ${R_{NL}\approx1.5}\,$k$\Omega$ at
${B=12}\,$T and ${T=300}\,$K, three times higher than that at
${T=10}\,$K.

\begin{figure}[t]
\centerline{\includegraphics[width=0.95\columnwidth]{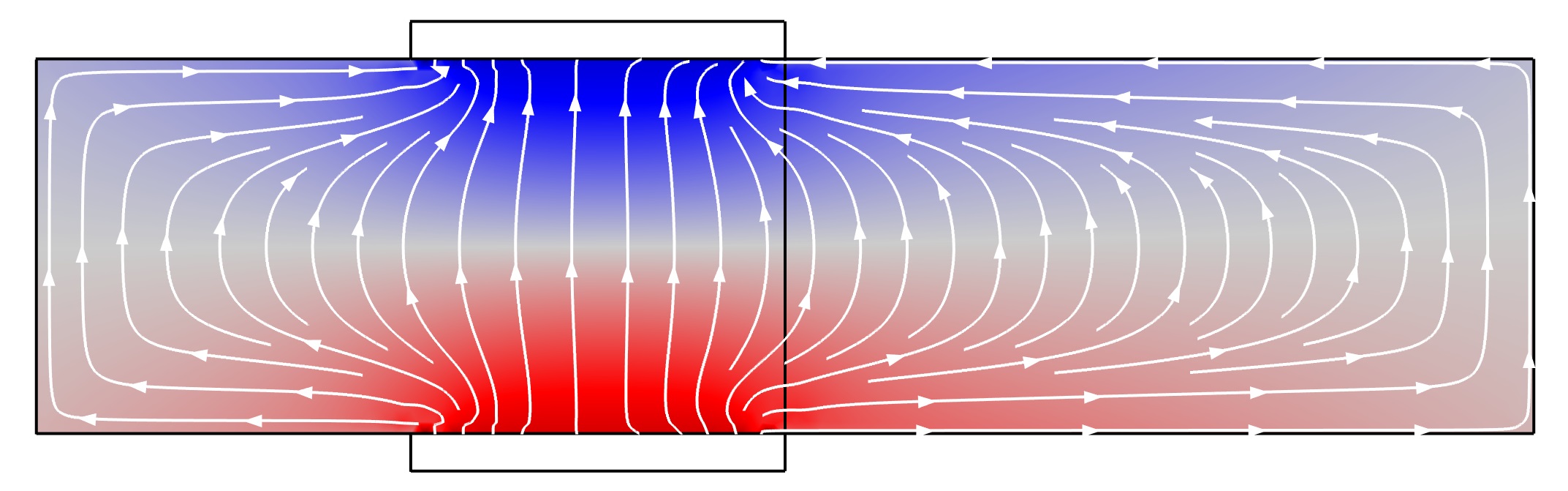}
}
\caption{Giant nonlocality in a compensated semimetal in magnetic
  field. The arrows indicate the current flow and the color map shows
  the electrochemical potential (see the main text and
  Figs.~\ref{fig2:sb} and \ref{fig3:tb} for specific parameters).}
\label{fig1:vortex}
\end{figure}

In this Letter, we argue that the giant nonlocality observed in
intrinsic graphene at high temperatures can be attributed to the
presence of two types of charge carriers (electrons and holes): at the
neutrality point, the two bands (the conductance and valence bands)
touch creating a two-component electronic system. Physics of such
systems is much richer than in their single-component
counterparts. Observed phenomena that are directly related to the
two-band structure of the neutrality point include giant magnetodrag
in graphene \cite{drag12,meg} and linear magnetoresistance
\cite{mr1,mrexp}. Both effects have been explained within a
phenomenological framework \cite{meg,mr1} allowing for a two-component
(electron-hole) system coupled by the external magnetic field. We
generalize this approach to investigate evolution of the spatial
distribution of the electron current density in the experimentally
relevant Hall bar geometry. In sufficiently strong magnetic fields,
the current density forms a giant nonlocal pattern where the current
is flowing not only in the bulk, but also along the boundaries leading
to strong nonlocal resistance, see Fig.~\ref{fig1:vortex}. Such
patterns can be directly observed in laboratory experiments using the
modern imaging techniques \cite{sulp,imh,imm}. Tuning the model
parameters to the specific values available for graphene, we arrive at
a quantitative estimate of the nonlocal resistance \cite{nlr}.

To highlight the difference between the one- and two-component
systems, we briefly recall the macroscopic description of electronic
transport in the standard (former) case. Allowing for nonuniform
charge density, the linear relation between the electric current
$\bs{J}$ and the external fields $\bs{E}$, $\bs{B}$ could be
formulated as \cite{dau10,df2,sven1}
\begin{subequations}
\label{sbeqs}  
\begin{equation}
\label{eq1}
r_0 \bs{J} = \bs{E} + r_H \bs{e}_{\bs{B}}\!\times\!\bs{J} + \frac{1}{e\nu_0} \bs{\nabla} n,
\end{equation}
where $e>0$ is the unit charge, $\nu_0$ is the density of states
(DoS), $n$ is the carrier density, ${\bs{e}_{\bs{B}}}$ is the unit
vector in the direction of the magnetic field, and $r_0$ and $r_H$ are
the longitudinal and Hall resistivities. Within the Drude-like
description, ${r_H=\omega_c\tau{r_0}}$ ($\omega_c$ is the cyclotron
frequency and $\tau$ is the mean free path). The relation
Eq.~(\ref{eq1}) is applicable to a wide range of electronic systems
from simple metals \cite{ziman,Giuliani} to doped graphene
\cite{kats,rev}. The transport coefficients $r_0$ and $r_H$ could be
treated as phenomenological or could be derived from the underlying
kinetic theory \cite{rev,me1,dau10}.

\begin{figure}[t]
\centerline{\includegraphics[width=0.95\columnwidth]{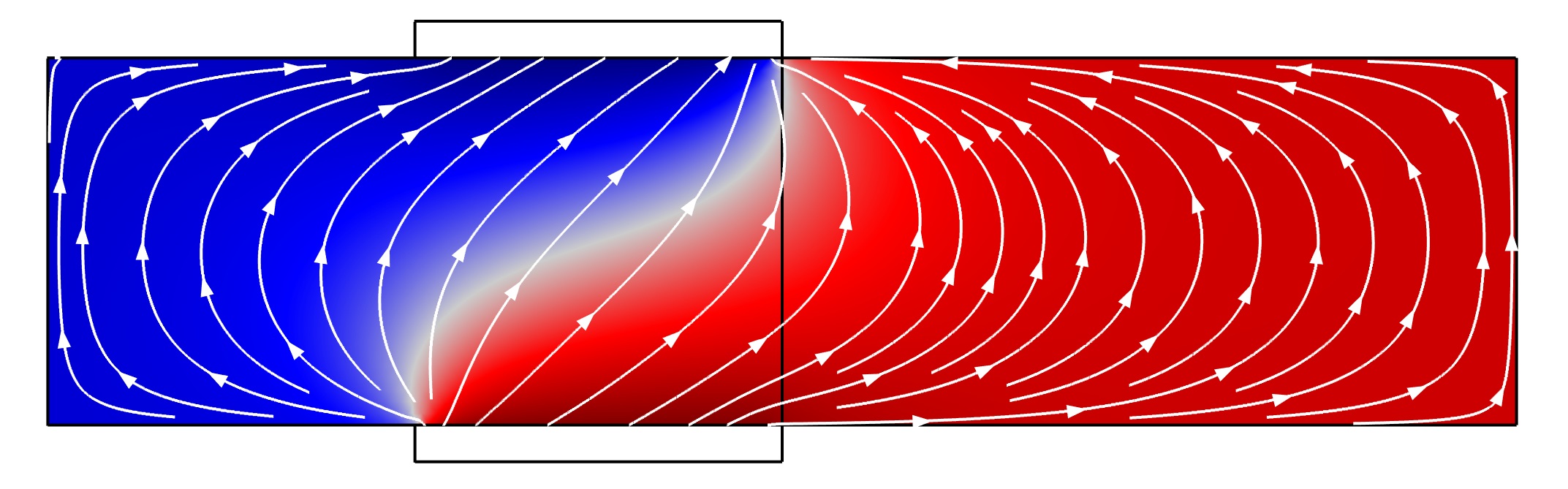}}
\caption{Classical Hall effect in a one-component electronic
  system. The current density (shown by the arrows) and the
  electrochemical potential (shown by the color map) were obtained
  from Eqs.~(\ref{sbeqs}) for a sample of the width $W=1\,\mu$m and
  length $L=4\,\mu$m with the carrier density $n=10^{12}\,$cm$^{-2}$
  at the temperature $T=240\,$K and in magnetic field $B=0.2\,$T.}
\label{fig2:sb}
\end{figure}

In addition to Eq.(\ref{eq1}), the electric current satisfies the
continuity equation, which for stationary currents reads
\begin{equation}
\label{ce}
\bs{\nabla}\!\cdot\!\bs{J}=0.
\end{equation}
Charge density inhomogeneity induces electric field, so that
Eq.~(\ref{eq1}) should be combined with the corresponding
electrostatic problem. Most recent experiments were performed in gated
structures, where the relation between the electric field and charge
density simplifies \cite{mr1,ash}. In two-dimensional (2D) samples
\begin{equation}
\label{vlas}
\bs{E} = \bs{E}_0 -\frac{e}{C}\bs{\nabla} n,
\end{equation}
\end{subequations}
where $C=\epsilon/(4\pi d)$ is the gate-to-sample capacitance per unit
area, $d$ is the distance to the gate, $\epsilon$ is the dielectric
constant, and $\bs{E}_0$ is the external field.

In a two-terminal (slab) geometry, solution of Eqs.~(\ref{sbeqs}) is a
textbook problem. In the absence of magnetic field, the resulting
electrochemical potential is governed by the relation of the mean free
path to the system size, exhibiting either a flat (in short, ballistic
samples) or linear (in long, diffusive samples) spatial profile. Most
recently, these solutions were used as benchmarks in the imaging
experiment \cite{sulp} and the numerical solution of the hydrodynamic
equations in doped graphene \cite{sven1}. In external magnetic field,
the system exhibits the classical Hall effect, which in short samples
is accompanied by nontrivial current flow patterns \cite{shik}.

In a four-terminal Hall bar geometry, the electric current still fills
the whole sample, but decays exponentially \cite{pauw} away from the
direct path between source and drain. The resulting flow pattern was
calculated (in the context of doped graphene) in
Refs.~\cite{fl0,fl1,sven1}. In magnetic field, the pattern gets skewed
due to the classical Hall effect, but exhibits no qualitatively new
features, see Fig.~\ref{fig2:sb}.

\begin{figure}[t]
\centerline{\includegraphics[width=0.95\columnwidth]{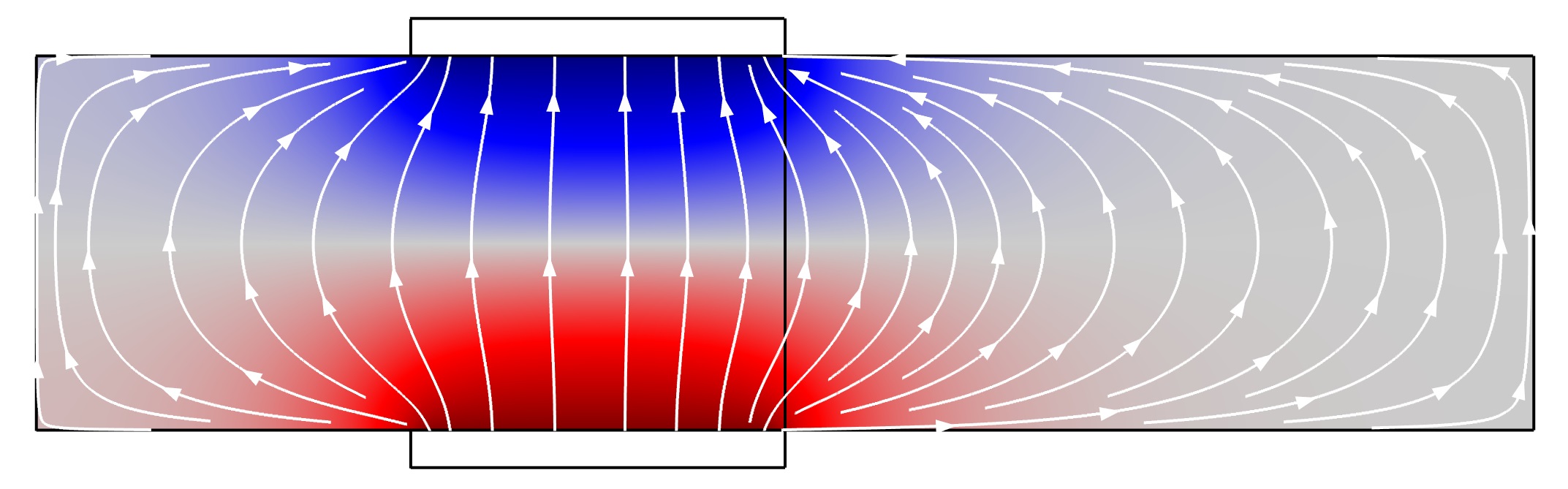}}
\centerline{\includegraphics[width=0.95\columnwidth]{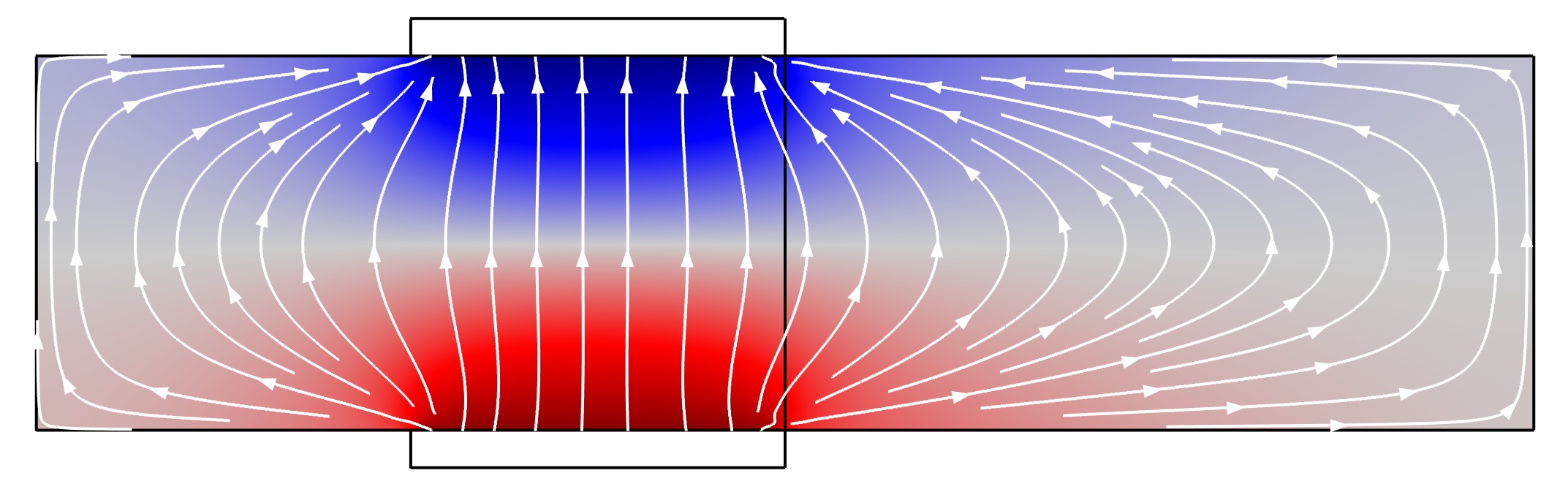}}
\caption{Charge flow in compensated semimetals. Top: Ohmic flow in the
  absence of magnetic field. Bottom: emergent nonlocality in weak
  magnetic field $B=0.2\,$T. The associated potential on the sample
  boundaries grows with the increasing field, see
  Fig.~\ref{fig1:vortex} for the pattern at $B=2\,$T. Stronger fields
  expel the current from the bulk such that it flows along the
  boundary.}
\label{fig3:tb}
\end{figure}

Let us now extend the transport equations (\ref{sbeqs}) to a
two-component system. Keeping in mind applications to graphene, we
re-write Eq.~(\ref{eq1}) for the quasiparticles in the conduction band
(``electrons'') in the form
\begin{subequations}
\label{eheqs}
\begin{equation}
\label{eqe}
-\bs{j}_e = e D \nu_e \bs{E} + \omega_c\tau \bs{j}_e\!\times\!\bs{e}_{\bs{B}} + D \bs{\nabla} n_e,
\end{equation}
where $\bs{j}_e$ is the electron flow density (carrying the electric
current ${\bs{J}_e=-e\bs{j}_e}$) and $\nu_e$ is DoS. The ``holes''
(i.e., the quasiparticles in the valence band) are described by
\begin{equation}
\label{eqh}
-\bs{j}_h = -e D \nu_h \bs{E} - \omega_c\tau \bs{j}_h\!\times\!\bs{e}_{\bs{B}} + D \bs{\nabla} n_h.
\end{equation}
\end{subequations}
Here the electric current carried by the holes is
${\bs{J}_h=e\bs{j}_h}$ and DoS may differ from that of electrons,
${\nu_h\ne\nu_e}$. For simplicity, we assume that the the cyclotron
frequency, mean free time, and diffusion constant for the two bands
coincide (a generalization is straightforward, but doesn't lead to
qualitatively new physics).

\begin{figure*}[t]
\centerline{\includegraphics[width=0.9\textwidth]{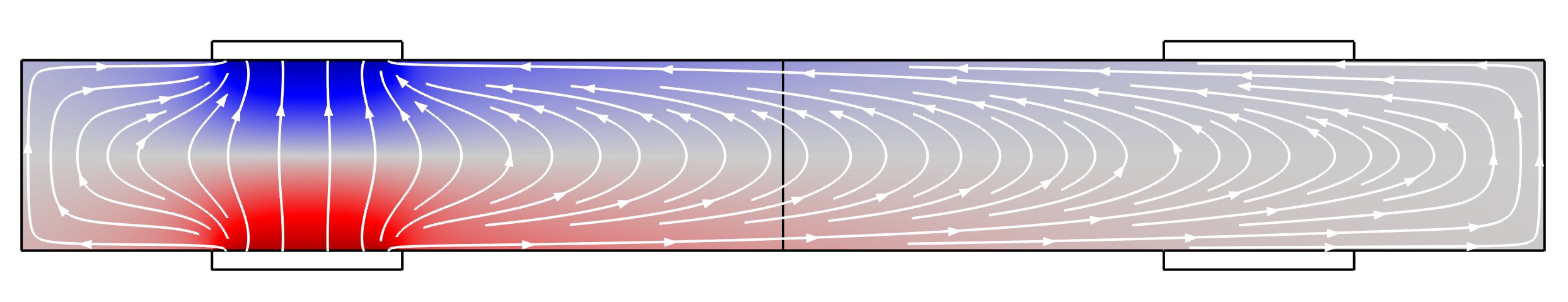}}
\caption{Giant nonlocality in the Hall bar geometry. The sample has a
  width ${W=1\,\mu}$m and length ${8\,\mu}$m, with the distance
  between contacts ${L=5\,\mu}$m. The driving current is
  ${I=0.1\,\mu}$A. The flow pattern was computed for ${B=0.8}\,$T,
  cf. Fig.~\ref{fig1:vortex}.}
\label{fig4:hall_bar}
\end{figure*}

The total electric current in the two component system is given by
${\bs{J}=-e\bs{j}}$, where ${\bs{j}=\bs{j}_e-\bs{j}_h}$. Introducing
also the total quasiparticle flow ${\bs{j}_I=\bs{j}_e+\bs{j}_h}$, we
find (cf. Ref.~\cite{me1})
\begin{subequations}
\label{eq2}
\begin{equation}
\label{eqj}
\bs{j}+eD(\nu_e+\nu_h)\bs{E}+\omega_c\tau\bs{j}_I\!\times\!\bs{e}_{\bs{B}} + D \bs{\nabla} n = 0,
\end{equation}
\begin{equation}
\label{eqp}
\bs{j}_I+eD(\nu_e-\nu_h)\bs{E}+\omega_c\tau\bs{j}\!\times\!\bs{e}_{\bs{B}} + D \bs{\nabla}\rho = 0,
\end{equation}
where ${n=n_e-n_h}$ is the carrier density per unit charge (the charge
density being $-en$) and ${\rho=n_e+n_h}$ is the total quasiparticle
density. The transport equations have to be supplemented by continuity
equations reflecting the particle number conservation. The electric
current satisfies Eq.~(\ref{ce}), but the total number of
quasiparticles \cite{alf} can be affected by electron-hole
recombination processes leading to a weak decay term in the continuity
equation
\begin{equation}
\label{cei}
\bs{\nabla}\!\cdot\!\bs{j}_I=-\delta\rho/\tau_R,
\end{equation}
\end{subequations}
where $\delta\rho$ is the deviation of the quasiparticle density from
its equilibrium value and $\tau_R$ is the recombination time.

Under the assumption of electron-hole symmetry (e.g., at the charge
neutrality point in graphene), ${\nu_e=\nu_h}$, we recover the
phenomenological model of Ref.~\cite{mr1}. In the two-terminal
geometry this model yields unsaturating linear magnetoresistance in
classically strong fields \cite{mrexp}.

Now we analyze the behavior of the phenomenological model (\ref{eq2})
in the four-terminal Hall bar geometry. In the absence of the magnetic
field, the system exhibits a typical Ohmic flow \cite{fl0,fl1,sven1},
see the top panel in Fig.~\ref{fig3:tb}. Applying the field we find a
qualitative change in the flow pattern -- the emergence of a boundary
flow and the associated electrochemical potential at the sample
edges. Increasing the field leads to the nonlocal pattern growing
until it fills the whole sample, see Figs.~\ref{fig1:vortex} and
\ref{fig4:hall_bar}.  Stronger fields essentially expel the current
from the bulk with the charge flow being concentrated along the sample
boundaries, which leads to strong nonlocal resistance.

The nonlocal flow pattern emerging in magnetic field, see
Figs.~\ref{fig1:vortex}, \ref{fig3:tb} and \ref{fig4:hall_bar}, has to
be contrasted with the vortices appearing in the viscous hydrodynamic
flow (e.g., in doped graphene \cite{fl0,fl1,sven1,lev19}). In the
latter case, vorticity appears due to the constrained geometry of the
flow and the particular boundary conditions \cite{fl1,sven1,ks19}:
neglecting Ohmic effects, the solution of the hydrodynamic equations
can be obtained by introducing the stream function, which obeys a
biharmonic equation independent of viscosity, which however affects
the distribution of the electrochemical potential. In contrast, within
the model (\ref{eq2}) the ``Ohmic'' scattering represents the only
source of dissipation and hence cannot be omitted. One can still
introduce the stream function, but now it is determined not only by
the sample geometry, but also by the Ohmic scattering and magnetic
field. As a result, the flow pattern does not exhibit vortices, unlike
those suggested recently for the hydrodynamic flow in intrinsic
graphene \cite{lev19} (in the absence of magnetic field).

Nonlocal resistance in graphene subjected to external magnetic field
was studied experimentally in Ref.~\cite{nlr}. At high enough
temperatures where signatures of the quantum Hall effect are washed
out, strong (or ``giant'') nonlocality was observed at the neutrality
point. The effect vanishes in zero field as well as with doping away
from neutrality. Both features are consistent with the model
(\ref{eq2}): in zero field the model exhibits usual Ohmic flow
patterns, see Fig.~\ref{fig3:tb}, while at sufficiently high doping
levels the effects of the second band are suppressed -- the two
equations (\ref{eqj}) and (\ref{eqp}) become identical showing the
response typical of one-component systems, see Fig.~\ref{fig2:sb}.

Having discussed the qualitative features of the charge flow in
two-component systems, we now turn to a quantitative calculation of
nonlocal resistance in graphene. Although the model (\ref{eq2}) is
applicable to any semimetal, graphene is a by far better studied
material with readily available experimental values for model
parameters. Here we use the data measured in
Refs.~\cite{nlr,gal,geim1,geim3,meg} and theoretical calculations of
Refs.~\cite{rev,luc,me1,meg,lev19}.

\begin{figure}[t]
\centerline{
\includegraphics[width=0.9\columnwidth]{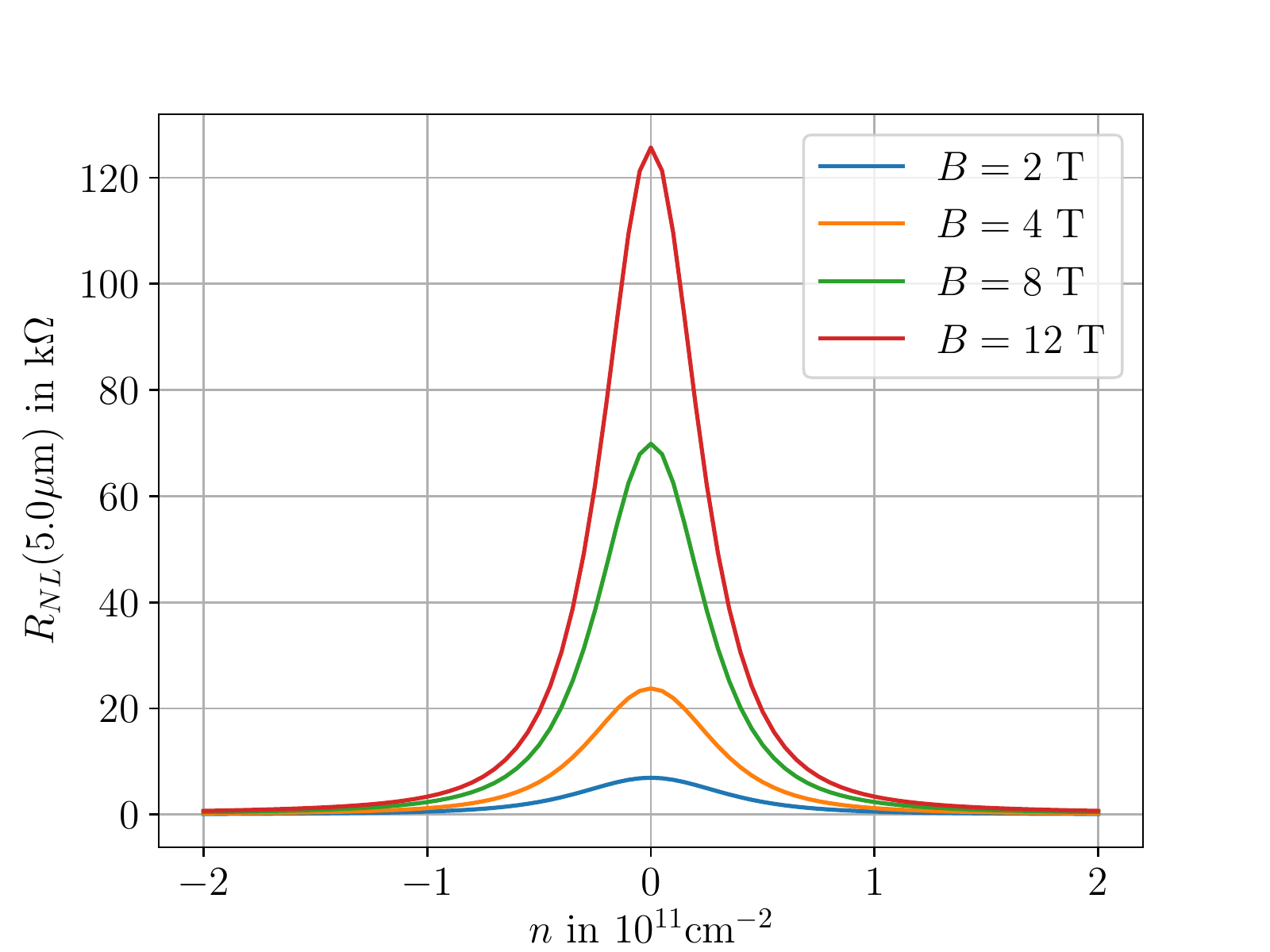}
}
\medskip
\centerline{
\includegraphics[width=0.9\columnwidth]{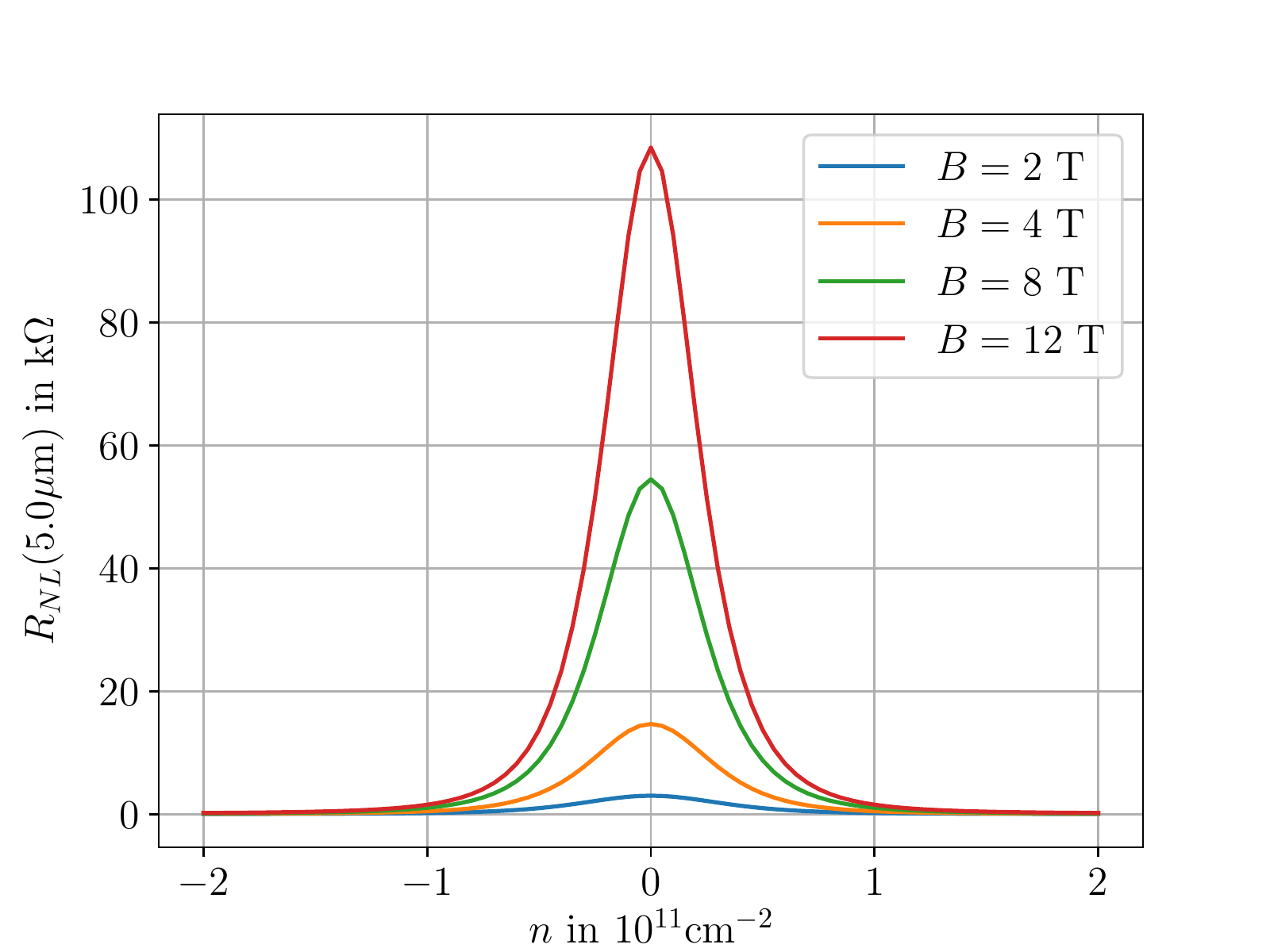}
}
\caption{Nonlocal resistance measured in the Hall bar geometry, see
  Fig.~\ref{fig4:hall_bar}, as a function of carrier density. Top:
  Coulomb scatterers; bottom: short-ranged impurities. The impurity
  model parameters are chosen to represent the mobility at
  ${n=10^{11}}\,$cm$^{-2}$ reported in Ref.~\cite{nlr}. The range of
  magnetic fields and carrier densities as well as the distance to the
  gate (${d=50}\,$nm) is taken from Ref.~\cite{nlr}, see Fig.2.}
\label{fig5:nlr}
\end{figure}

DoS of the quasiparticles in graphene has been evaluated in, e.g.,
Refs.~\cite{rev,luc,me1,kats}, and has the form
\begin{equation}
\label{dos}
\nu_e+\nu_h = 2{\cal T}/(\pi v_g^2), 
\qquad
\nu_e-\nu_h = 2\mu/(\pi v_g^2), 
\end{equation}
where $\mu$ is the chemical potential, $v_g$ is the quasiparticle
velocity in graphene, and ${{\cal{T}}=2T\ln[2\cosh(\mu/2T)]}$. The
generalized cyclotron frequency is ${\omega_c=eBv_g^2/(c{\cal{T}})}$
and the diffusion coefficient has the usual form ${D=v_g^2\tau/2}$. At
charge neutrality, ${\mu=0}$ and ${{\cal{T}}=2T\ln2}$, while in the
degenerate regime ${{\cal{T}}(\mu\gg{T})=\mu}$. The latter confirms
that all coefficients in Eqs.~(\ref{eqj}) and (\ref{eqp}) become
identical with doping. Similarly, the continuity equations (\ref{ce})
and (\ref{cei}) should coincide in the degenerate regime. In graphene
this happens by means of the fast decay of the recombination rate
\cite{meg}. Close to neutrality we assume
\begin{equation}
\label{tr}
\tau_R^{-1} = g^2T/\cosh(\mu/T),
\end{equation}
where $g$ is determined by the corresponding matrix element. The above
expression \cite{meg} reflects the exponential decay of the two-band
physics away from charge neutrality, which is responsible for the fast
decay of $R_{NL}$ as a function of carrier density \cite{nlr}, see
Fig.~\ref{fig5:nlr}. Finally, the mean-free time, $\tau$, in graphene
is a non-trivial function of temperature and carrier density
\cite{kats,rev,luc,gal,drag}, which strongly depends on the model of
the impurity potential \cite{ando,ando2,nom,fal,alef,ogm}. However,
these dependencies are typically not exponential and hence do not
affect the exponential decay of the nonlocal resistance.

In Fig.~\ref{fig5:nlr} we demonstrate the decay of $R_{NL}$ for two
impurity models -- the Coulomb scatterers and short-ranged impurities
-- showing nearly identical behavior. Such robustness of the model
(\ref{eq2}) with respect of the functional dependence of the mean free
time justifies the inaccuracy of our description of electronic
transport in graphene, where close to charge neutrality the
resistivity is strongly affected by electron-electron interaction. The
data shown in Fig.~\ref{fig5:nlr} were obtained by solving
Eqs.~(\ref{eq2}) in the Hall bar geometry of Fig.~\ref{fig4:hall_bar}
using the estimate \cite{lev19} for the recombination length scale,
${\ell_R=v_g\tau_R\approx10\,\mu}$m (a previous calculation of
Ref.~\cite{meg} put it at a smaller value $1.2\,\mu$m), which leads to
similar results for the nonlocal resistance, but with a smaller peak
value at charge neutrality.

The results for $R_{NL}$ shown in Fig.~\ref{fig5:nlr} are extremely
similar to those reported in Ref.~\cite{nlr} with the exception of the
values at neutrality, which are grossly exaggerated. There are several
reasons for this behavior. Firstly, by ignoring the effects of
electron-electron interaction, we strongly underestimate the usual
resistivity of intrinsic graphene. Secondly, we ignore viscous
effects. Furthermore, DoS in real graphene never really vanishes ``at
neutrality'' due to electrostatic potential fluctuations
\cite{chiap}. As a result, the minimal carrier concentration is often
as high as $10^{10}$cm$^{-2}$, essentially cutting off the lower
density range around the peak in Fig.~\ref{fig5:nlr}. Finally,
Eq.~(\ref{tr}) is a rather crude estimate that needs to be improved.

To conclude, we have argued that the observed giant nonlocality in
neutral graphene in non-quantizing magnetic fields at relatively high
temperatures observed in Ref.~\cite{nlr} is a direct consequence of
the two-band nature of the quasiparticle spectrum in graphene. As
such, this effect is not specific to graphene and should be observable
in any compensated two-component system. Our theory does not involve
spin-related phenomena including the effect of Zeeman splitting
invoked in Ref.~\cite{nlr}. The latter should be independent of the
field direction, however, the effect was not observed in the nearly
parallel field studied in Ref.~\cite{chiap}. Assuming the $g$-factor
to be equal to $2$, we estimate the Zeeman splitting as
${E_z\approx0.35}\,$meV${\approx4}\,$K at ${B=10}\,$T. The
corresponding residual quasiparticle density (at ${T=0}$) is given by
${\rho_Q=E_z^2/(4\pi v_g^2)\approx2.2\times10^6}\,$cm$^{-2}$. As a
result, we expect the effects of Zeeman splitting to be observable at
temperatures and carrier densities much lower than those typical to
nonlocal measurements discussed here.

With material-specific parameters, our phenomenological model is
capable of a quantitative description of the effect. For graphene, a
more precise calculation involving solution of the full system of
hydrodynamic equations near charge neutrality is required to reach
perfect agreement with the data, however the present approach shows
that the effect is more general and does not require additional
assumptions of electronic hydrodynamics.


The authors are grateful to I.V. Gornyi, A.D. Mirlin, J. Schmalian,
J.A. Sulpizio, M. Sch\"utt, A. Shnirman, and Y. Tserkovnyak for
fruitful discussions. This work was supported by the German Research
Foundation DFG within FLAG-ERA Joint Transnational Call (Project
GRANSPORT), by the European Commission under the EU Horizon 2020
MSCA-RISE-2019 program (Project 873028 HYDROTRONICS), and by the
Russian Science Foundation Project No. 17-12-0 (MT). BNN acknowledges
the support by the MEPhI Academic Excellence Project, Contract
No. 02.a03.21.0005.



\bibliography{viscosity_refs}

\begin{thebibliography}{51}
\expandafter\ifx\csname natexlab\endcsname\relax\def\natexlab#1{#1}\fi
\expandafter\ifx\csname bibnamefont\endcsname\relax
  \def\bibnamefont#1{#1}\fi
\expandafter\ifx\csname bibfnamefont\endcsname\relax
  \def\bibfnamefont#1{#1}\fi
\expandafter\ifx\csname citenamefont\endcsname\relax
  \def\citenamefont#1{#1}\fi
\expandafter\ifx\csname url\endcsname\relax
  \def\url#1{\texttt{#1}}\fi
\expandafter\ifx\csname urlprefix\endcsname\relax\def\urlprefix{URL }\fi
\providecommand{\bibinfo}[2]{#2}
\providecommand{\eprint}[2][]{\url{#2}}

\bibitem[{\citenamefont{Skocpol et~al.}(1987)\citenamefont{Skocpol, Mankiewich,
  Howard, Jackel, Tennant, and Stone}}]{skocpol}
\bibinfo{author}{\bibfnamefont{W.~J.} \bibnamefont{Skocpol}},
  \bibinfo{author}{\bibfnamefont{P.~M.} \bibnamefont{Mankiewich}},
  \bibinfo{author}{\bibfnamefont{R.~E.} \bibnamefont{Howard}},
  \bibinfo{author}{\bibfnamefont{L.~D.} \bibnamefont{Jackel}},
  \bibinfo{author}{\bibfnamefont{D.~M.} \bibnamefont{Tennant}},
  \bibnamefont{and} \bibinfo{author}{\bibfnamefont{A.~D.} \bibnamefont{Stone}},
  \bibinfo{journal}{Phys. Rev. Lett.} \textbf{\bibinfo{volume}{58}},
  \bibinfo{pages}{2347} (\bibinfo{year}{1987}).

\bibitem[{\citenamefont{van Houten et~al.}(1989)\citenamefont{van Houten,
  Beenakker, Williamson, Broekaart, van Loosdrecht, van Wees, Mooij, Foxon, and
  Harris}}]{vwees1}
\bibinfo{author}{\bibfnamefont{H.}~\bibnamefont{van Houten}},
  \bibinfo{author}{\bibfnamefont{C.~W.~J.} \bibnamefont{Beenakker}},
  \bibinfo{author}{\bibfnamefont{J.~G.} \bibnamefont{Williamson}},
  \bibinfo{author}{\bibfnamefont{M.~E.~I.} \bibnamefont{Broekaart}},
  \bibinfo{author}{\bibfnamefont{P.~H.~M.} \bibnamefont{van Loosdrecht}},
  \bibinfo{author}{\bibfnamefont{B.~J.} \bibnamefont{van Wees}},
  \bibinfo{author}{\bibfnamefont{J.~E.} \bibnamefont{Mooij}},
  \bibinfo{author}{\bibfnamefont{C.~T.} \bibnamefont{Foxon}}, \bibnamefont{and}
  \bibinfo{author}{\bibfnamefont{J.~J.} \bibnamefont{Harris}},
  \bibinfo{journal}{Phys. Rev. B} \textbf{\bibinfo{volume}{39}},
  \bibinfo{pages}{8556} (\bibinfo{year}{1989}).

\bibitem[{\citenamefont{Geim et~al.}(1991)\citenamefont{Geim, Main, Beton,
  Streda, Eaves, Wilkinson, and Beaumont}}]{geimnl1}
\bibinfo{author}{\bibfnamefont{A.~K.} \bibnamefont{Geim}},
  \bibinfo{author}{\bibfnamefont{P.~C.} \bibnamefont{Main}},
  \bibinfo{author}{\bibfnamefont{P.~H.} \bibnamefont{Beton}},
  \bibinfo{author}{\bibfnamefont{P.}~\bibnamefont{Streda}},
  \bibinfo{author}{\bibfnamefont{L.}~\bibnamefont{Eaves}},
  \bibinfo{author}{\bibfnamefont{C.~D.~W.} \bibnamefont{Wilkinson}},
  \bibnamefont{and} \bibinfo{author}{\bibfnamefont{S.~P.}
  \bibnamefont{Beaumont}}, \bibinfo{journal}{Phys. Rev. Lett.}
  \textbf{\bibinfo{volume}{67}}, \bibinfo{pages}{3014} (\bibinfo{year}{1991}).

\bibitem[{\citenamefont{Shepard et~al.}(1992)\citenamefont{Shepard, Roukes, and
  Van~der Gaag}}]{roukes}
\bibinfo{author}{\bibfnamefont{K.~L.} \bibnamefont{Shepard}},
  \bibinfo{author}{\bibfnamefont{M.~L.} \bibnamefont{Roukes}},
  \bibnamefont{and} \bibinfo{author}{\bibfnamefont{B.~P.} \bibnamefont{Van~der
  Gaag}}, \bibinfo{journal}{Phys. Rev. Lett.} \textbf{\bibinfo{volume}{68}},
  \bibinfo{pages}{2660} (\bibinfo{year}{1992}).

\bibitem[{\citenamefont{Hirayama et~al.}(1992)\citenamefont{Hirayama, Wieck,
  Bever, von Klitzing, and Ploog}}]{vonklit}
\bibinfo{author}{\bibfnamefont{Y.}~\bibnamefont{Hirayama}},
  \bibinfo{author}{\bibfnamefont{A.~D.} \bibnamefont{Wieck}},
  \bibinfo{author}{\bibfnamefont{T.}~\bibnamefont{Bever}},
  \bibinfo{author}{\bibfnamefont{K.}~\bibnamefont{von Klitzing}},
  \bibnamefont{and} \bibinfo{author}{\bibfnamefont{K.}~\bibnamefont{Ploog}},
  \bibinfo{journal}{Phys. Rev. B} \textbf{\bibinfo{volume}{46}},
  \bibinfo{pages}{4035} (\bibinfo{year}{1992}).

\bibitem[{\citenamefont{Mihajlovi\ifmmode~\acute{c}\else \'{c}\fi{}
  et~al.}(2009)\citenamefont{Mihajlovi\ifmmode~\acute{c}\else \'{c}\fi{},
  Pearson, Garcia, Bader, and Hoffmann}}]{nlgold}
\bibinfo{author}{\bibfnamefont{G.}~\bibnamefont{Mihajlovi\ifmmode~\acute{c}\else
  \'{c}\fi{}}}, \bibinfo{author}{\bibfnamefont{J.~E.} \bibnamefont{Pearson}},
  \bibinfo{author}{\bibfnamefont{M.~A.} \bibnamefont{Garcia}},
  \bibinfo{author}{\bibfnamefont{S.~D.} \bibnamefont{Bader}}, \bibnamefont{and}
  \bibinfo{author}{\bibfnamefont{A.}~\bibnamefont{Hoffmann}},
  \bibinfo{journal}{Phys. Rev. Lett.} \textbf{\bibinfo{volume}{103}},
  \bibinfo{pages}{166601} (\bibinfo{year}{2009}).

\bibitem[{\citenamefont{Gorbachev et~al.}(2014)\citenamefont{Gorbachev, Song,
  Yu, Kretinin, Withers, Cao, Mishchenko, Grigorieva, Novoselov, Levitov
  et~al.}}]{geimnl2}
\bibinfo{author}{\bibfnamefont{R.~V.} \bibnamefont{Gorbachev}},
  \bibinfo{author}{\bibfnamefont{J.~C.~W.} \bibnamefont{Song}},
  \bibinfo{author}{\bibfnamefont{G.~L.} \bibnamefont{Yu}},
  \bibinfo{author}{\bibfnamefont{A.~V.} \bibnamefont{Kretinin}},
  \bibinfo{author}{\bibfnamefont{F.}~\bibnamefont{Withers}},
  \bibinfo{author}{\bibfnamefont{Y.}~\bibnamefont{Cao}},
  \bibinfo{author}{\bibfnamefont{A.}~\bibnamefont{Mishchenko}},
  \bibinfo{author}{\bibfnamefont{I.~V.} \bibnamefont{Grigorieva}},
  \bibinfo{author}{\bibfnamefont{K.~S.} \bibnamefont{Novoselov}},
  \bibinfo{author}{\bibfnamefont{L.~S.} \bibnamefont{Levitov}},
  \bibnamefont{et~al.}, \bibinfo{journal}{Science}
  \textbf{\bibinfo{volume}{346}}, \bibinfo{pages}{448} (\bibinfo{year}{2014}).

\bibitem[{\citenamefont{Bandurin et~al.}(2016)\citenamefont{Bandurin, Torre,
  Krishna~Kumar, Ben~Shalom, Tomadin, Principi, Auton, Khestanova, Novoselov,
  Grigorieva et~al.}}]{geim1}
\bibinfo{author}{\bibfnamefont{D.~A.} \bibnamefont{Bandurin}},
  \bibinfo{author}{\bibfnamefont{I.}~\bibnamefont{Torre}},
  \bibinfo{author}{\bibfnamefont{R.}~\bibnamefont{Krishna~Kumar}},
  \bibinfo{author}{\bibfnamefont{M.}~\bibnamefont{Ben~Shalom}},
  \bibinfo{author}{\bibfnamefont{A.}~\bibnamefont{Tomadin}},
  \bibinfo{author}{\bibfnamefont{A.}~\bibnamefont{Principi}},
  \bibinfo{author}{\bibfnamefont{G.~H.} \bibnamefont{Auton}},
  \bibinfo{author}{\bibfnamefont{E.}~\bibnamefont{Khestanova}},
  \bibinfo{author}{\bibfnamefont{K.~S.} \bibnamefont{Novoselov}},
  \bibinfo{author}{\bibfnamefont{I.~V.} \bibnamefont{Grigorieva}},
  \bibnamefont{et~al.}, \bibinfo{journal}{Science}
  \textbf{\bibinfo{volume}{351}}, \bibinfo{pages}{1055} (\bibinfo{year}{2016}).

\bibitem[{\citenamefont{Bandurin et~al.}(2018)\citenamefont{Bandurin, Shytov,
  Levitov, Kumar, Berdyugin, {Ben Shalom}, Grigorieva, Geim, and
  Falkovich}}]{geim3}
\bibinfo{author}{\bibfnamefont{D.~A.} \bibnamefont{Bandurin}},
  \bibinfo{author}{\bibfnamefont{A.~V.} \bibnamefont{Shytov}},
  \bibinfo{author}{\bibfnamefont{L.~S.} \bibnamefont{Levitov}},
  \bibinfo{author}{\bibfnamefont{R.~K.} \bibnamefont{Kumar}},
  \bibinfo{author}{\bibfnamefont{A.~I.} \bibnamefont{Berdyugin}},
  \bibinfo{author}{\bibfnamefont{M.}~\bibnamefont{{Ben Shalom}}},
  \bibinfo{author}{\bibfnamefont{I.~V.} \bibnamefont{Grigorieva}},
  \bibinfo{author}{\bibfnamefont{A.~K.} \bibnamefont{Geim}}, \bibnamefont{and}
  \bibinfo{author}{\bibfnamefont{G.}~\bibnamefont{Falkovich}},
  \bibinfo{journal}{Nat. Commun.} \textbf{\bibinfo{volume}{9}},
  \bibinfo{pages}{4533} (\bibinfo{year}{2018}).

\bibitem[{\citenamefont{Berdyugin et~al.}(2019)\citenamefont{Berdyugin, Xu,
  Pellegrino, Kumar, Principi, Torre, Shalom, Taniguchi, Watanabe, Grigorieva
  et~al.}}]{geim4}
\bibinfo{author}{\bibfnamefont{A.~I.} \bibnamefont{Berdyugin}},
  \bibinfo{author}{\bibfnamefont{S.~G.} \bibnamefont{Xu}},
  \bibinfo{author}{\bibfnamefont{F.~M.~D.} \bibnamefont{Pellegrino}},
  \bibinfo{author}{\bibfnamefont{R.~K.} \bibnamefont{Kumar}},
  \bibinfo{author}{\bibfnamefont{A.}~\bibnamefont{Principi}},
  \bibinfo{author}{\bibfnamefont{I.}~\bibnamefont{Torre}},
  \bibinfo{author}{\bibfnamefont{M.~B.} \bibnamefont{Shalom}},
  \bibinfo{author}{\bibfnamefont{T.}~\bibnamefont{Taniguchi}},
  \bibinfo{author}{\bibfnamefont{K.}~\bibnamefont{Watanabe}},
  \bibinfo{author}{\bibfnamefont{I.~V.} \bibnamefont{Grigorieva}},
  \bibnamefont{et~al.}, \bibinfo{journal}{Science}
  \textbf{\bibinfo{volume}{364}}, \bibinfo{pages}{162} (\bibinfo{year}{2019}).

\bibitem[{\citenamefont{Narozhny et~al.}(2017)\citenamefont{Narozhny, Gornyi,
  Mirlin, and Schmalian}}]{rev}
\bibinfo{author}{\bibfnamefont{B.~N.} \bibnamefont{Narozhny}},
  \bibinfo{author}{\bibfnamefont{I.~V.} \bibnamefont{Gornyi}},
  \bibinfo{author}{\bibfnamefont{A.~D.} \bibnamefont{Mirlin}},
  \bibnamefont{and}
  \bibinfo{author}{\bibfnamefont{J.}~\bibnamefont{Schmalian}},
  \bibinfo{journal}{Annalen der Physik} \textbf{\bibinfo{volume}{529}},
  \bibinfo{pages}{1700043} (\bibinfo{year}{2017}).

\bibitem[{\citenamefont{Lucas and Fong}(2018)}]{luc}
\bibinfo{author}{\bibfnamefont{A.}~\bibnamefont{Lucas}} \bibnamefont{and}
  \bibinfo{author}{\bibfnamefont{K.~C.} \bibnamefont{Fong}},
  \bibinfo{journal}{J. Phys: Condens. Matter} \textbf{\bibinfo{volume}{30}},
  \bibinfo{pages}{053001} (\bibinfo{year}{2018}).

\bibitem[{\citenamefont{Torre et~al.}(2015)\citenamefont{Torre, Tomadin, Geim,
  and Polini}}]{pol15}
\bibinfo{author}{\bibfnamefont{I.}~\bibnamefont{Torre}},
  \bibinfo{author}{\bibfnamefont{A.}~\bibnamefont{Tomadin}},
  \bibinfo{author}{\bibfnamefont{A.~K.} \bibnamefont{Geim}}, \bibnamefont{and}
  \bibinfo{author}{\bibfnamefont{M.}~\bibnamefont{Polini}},
  \bibinfo{journal}{Phys. Rev. B} \textbf{\bibinfo{volume}{92}},
  \bibinfo{pages}{165433} (\bibinfo{year}{2015}).

\bibitem[{\citenamefont{Levitov and Falkovich}(2016)}]{fl0}
\bibinfo{author}{\bibfnamefont{L.}~\bibnamefont{Levitov}} \bibnamefont{and}
  \bibinfo{author}{\bibfnamefont{G.}~\bibnamefont{Falkovich}},
  \bibinfo{journal}{Nat. Phys.} \textbf{\bibinfo{volume}{12}},
  \bibinfo{pages}{672} (\bibinfo{year}{2016}).

\bibitem[{\citenamefont{Falkovich and Levitov}(2017)}]{fl1}
\bibinfo{author}{\bibfnamefont{G.}~\bibnamefont{Falkovich}} \bibnamefont{and}
  \bibinfo{author}{\bibfnamefont{L.}~\bibnamefont{Levitov}},
  \bibinfo{journal}{Phys. Rev. Lett.} \textbf{\bibinfo{volume}{119}},
  \bibinfo{pages}{066601} (\bibinfo{year}{2017}).

\bibitem[{\citenamefont{Pellegrino et~al.}(2017)\citenamefont{Pellegrino,
  Torre, and Polini}}]{pol17}
\bibinfo{author}{\bibfnamefont{F.~M.~D.} \bibnamefont{Pellegrino}},
  \bibinfo{author}{\bibfnamefont{I.}~\bibnamefont{Torre}}, \bibnamefont{and}
  \bibinfo{author}{\bibfnamefont{M.}~\bibnamefont{Polini}},
  \bibinfo{journal}{Phys. Rev. B} \textbf{\bibinfo{volume}{96}},
  \bibinfo{pages}{195401} (\bibinfo{year}{2017}).

\bibitem[{\citenamefont{Danz and Narozhny}(2019)}]{sven1}
\bibinfo{author}{\bibfnamefont{S.}~\bibnamefont{Danz}} \bibnamefont{and}
  \bibinfo{author}{\bibfnamefont{B.~N.} \bibnamefont{Narozhny}}
  (\bibinfo{year}{2019}), \eprint{arXiv:1910.14473}.

\bibitem[{\citenamefont{McEuen et~al.}(1990)\citenamefont{McEuen, Szafer,
  Richter, Alphenaar, Jain, Stone, Wheeler, and Sacks}}]{mceu}
\bibinfo{author}{\bibfnamefont{P.~L.} \bibnamefont{McEuen}},
  \bibinfo{author}{\bibfnamefont{A.}~\bibnamefont{Szafer}},
  \bibinfo{author}{\bibfnamefont{C.~A.} \bibnamefont{Richter}},
  \bibinfo{author}{\bibfnamefont{B.~W.} \bibnamefont{Alphenaar}},
  \bibinfo{author}{\bibfnamefont{J.~K.} \bibnamefont{Jain}},
  \bibinfo{author}{\bibfnamefont{A.~D.} \bibnamefont{Stone}},
  \bibinfo{author}{\bibfnamefont{R.~G.} \bibnamefont{Wheeler}},
  \bibnamefont{and} \bibinfo{author}{\bibfnamefont{R.~N.} \bibnamefont{Sacks}},
  \bibinfo{journal}{Phys. Rev. Lett.} \textbf{\bibinfo{volume}{64}},
  \bibinfo{pages}{2062} (\bibinfo{year}{1990}).

\bibitem[{\citenamefont{Wang and Goldman}(1992)}]{goldman}
\bibinfo{author}{\bibfnamefont{J.~K.} \bibnamefont{Wang}} \bibnamefont{and}
  \bibinfo{author}{\bibfnamefont{V.~J.} \bibnamefont{Goldman}},
  \bibinfo{journal}{Phys. Rev. B} \textbf{\bibinfo{volume}{45}},
  \bibinfo{pages}{13479} (\bibinfo{year}{1992}).

\bibitem[{\citenamefont{Roth et~al.}(2009)\citenamefont{Roth, Brune, Buhmann,
  Molenkamp, Maciejko, Qi, and Zhang}}]{roth}
\bibinfo{author}{\bibfnamefont{A.}~\bibnamefont{Roth}},
  \bibinfo{author}{\bibfnamefont{C.}~\bibnamefont{Brune}},
  \bibinfo{author}{\bibfnamefont{H.}~\bibnamefont{Buhmann}},
  \bibinfo{author}{\bibfnamefont{L.~W.} \bibnamefont{Molenkamp}},
  \bibinfo{author}{\bibfnamefont{J.}~\bibnamefont{Maciejko}},
  \bibinfo{author}{\bibfnamefont{X.-L.} \bibnamefont{Qi}}, \bibnamefont{and}
  \bibinfo{author}{\bibfnamefont{S.-C.} \bibnamefont{Zhang}},
  \bibinfo{journal}{Science} \textbf{\bibinfo{volume}{325}},
  \bibinfo{pages}{294} (\bibinfo{year}{2009}).

\bibitem[{\citenamefont{Abanin et~al.}(2011)\citenamefont{Abanin, Morozov,
  Ponomarenko, Gorbachev, Mayorov, Katsnelson, Watanabe, Taniguchi, Novoselov,
  Levitov et~al.}}]{nlr}
\bibinfo{author}{\bibfnamefont{D.~A.} \bibnamefont{Abanin}},
  \bibinfo{author}{\bibfnamefont{S.~V.} \bibnamefont{Morozov}},
  \bibinfo{author}{\bibfnamefont{L.~A.} \bibnamefont{Ponomarenko}},
  \bibinfo{author}{\bibfnamefont{R.~V.} \bibnamefont{Gorbachev}},
  \bibinfo{author}{\bibfnamefont{A.~S.} \bibnamefont{Mayorov}},
  \bibinfo{author}{\bibfnamefont{M.~I.} \bibnamefont{Katsnelson}},
  \bibinfo{author}{\bibfnamefont{K.}~\bibnamefont{Watanabe}},
  \bibinfo{author}{\bibfnamefont{T.}~\bibnamefont{Taniguchi}},
  \bibinfo{author}{\bibfnamefont{K.~S.} \bibnamefont{Novoselov}},
  \bibinfo{author}{\bibfnamefont{L.~S.} \bibnamefont{Levitov}},
  \bibnamefont{et~al.}, \bibinfo{journal}{Science}
  \textbf{\bibinfo{volume}{332}}, \bibinfo{pages}{328} (\bibinfo{year}{2011}).

\bibitem[{\citenamefont{Zhang et~al.}(2017)\citenamefont{Zhang, Huang, and
  Cazalilla}}]{caza}
\bibinfo{author}{\bibfnamefont{X.-P.} \bibnamefont{Zhang}},
  \bibinfo{author}{\bibfnamefont{C.}~\bibnamefont{Huang}}, \bibnamefont{and}
  \bibinfo{author}{\bibfnamefont{M.~A.} \bibnamefont{Cazalilla}},
  \bibinfo{journal}{2D Materials} \textbf{\bibinfo{volume}{4}},
  \bibinfo{pages}{024007} (\bibinfo{year}{2017}).

\bibitem[{\citenamefont{Komatsu et~al.}(2018)\citenamefont{Komatsu, Morita,
  Watanabe, Tsuya, Watanabe, Taniguchi, and Moriyama}}]{koma}
\bibinfo{author}{\bibfnamefont{K.}~\bibnamefont{Komatsu}},
  \bibinfo{author}{\bibfnamefont{Y.}~\bibnamefont{Morita}},
  \bibinfo{author}{\bibfnamefont{E.}~\bibnamefont{Watanabe}},
  \bibinfo{author}{\bibfnamefont{D.}~\bibnamefont{Tsuya}},
  \bibinfo{author}{\bibfnamefont{K.}~\bibnamefont{Watanabe}},
  \bibinfo{author}{\bibfnamefont{T.}~\bibnamefont{Taniguchi}},
  \bibnamefont{and} \bibinfo{author}{\bibfnamefont{S.}~\bibnamefont{Moriyama}},
  \bibinfo{journal}{Science Advances} \textbf{\bibinfo{volume}{4}},
  \bibinfo{pages}{eaaq0194} (\bibinfo{year}{2018}).

\bibitem[{\citenamefont{van~der Pauw}(1958)}]{pauw}
\bibinfo{author}{\bibfnamefont{L.~J.} \bibnamefont{van~der Pauw}},
  \bibinfo{journal}{Philips Tech. Rev.} \textbf{\bibinfo{volume}{20}},
  \bibinfo{pages}{223} (\bibinfo{year}{1958}).

\bibitem[{\citenamefont{Gorbachev et~al.}(2012)\citenamefont{Gorbachev, Geim,
  Katsnelson, Novoselov, Tudorovskiy, Grigorieva, MacDonald, Morozov, Watanabe,
  Taniguchi et~al.}}]{drag12}
\bibinfo{author}{\bibfnamefont{R.~V.} \bibnamefont{Gorbachev}},
  \bibinfo{author}{\bibfnamefont{A.~K.} \bibnamefont{Geim}},
  \bibinfo{author}{\bibfnamefont{M.~I.} \bibnamefont{Katsnelson}},
  \bibinfo{author}{\bibfnamefont{K.~S.} \bibnamefont{Novoselov}},
  \bibinfo{author}{\bibfnamefont{T.}~\bibnamefont{Tudorovskiy}},
  \bibinfo{author}{\bibfnamefont{I.~V.} \bibnamefont{Grigorieva}},
  \bibinfo{author}{\bibfnamefont{A.~H.} \bibnamefont{MacDonald}},
  \bibinfo{author}{\bibfnamefont{S.~V.} \bibnamefont{Morozov}},
  \bibinfo{author}{\bibfnamefont{K.}~\bibnamefont{Watanabe}},
  \bibinfo{author}{\bibfnamefont{T.}~\bibnamefont{Taniguchi}},
  \bibnamefont{et~al.}, \bibinfo{journal}{Nat. Phys.}
  \textbf{\bibinfo{volume}{8}}, \bibinfo{pages}{896} (\bibinfo{year}{2012}).

\bibitem[{\citenamefont{Titov et~al.}(2013)\citenamefont{Titov, Gorbachev,
  Narozhny, Tudorovskiy, Sch\"utt, Ostrovsky, Gornyi, Mirlin, Katsnelson,
  Novoselov et~al.}}]{meg}
\bibinfo{author}{\bibfnamefont{M.}~\bibnamefont{Titov}},
  \bibinfo{author}{\bibfnamefont{R.~V.} \bibnamefont{Gorbachev}},
  \bibinfo{author}{\bibfnamefont{B.~N.} \bibnamefont{Narozhny}},
  \bibinfo{author}{\bibfnamefont{T.}~\bibnamefont{Tudorovskiy}},
  \bibinfo{author}{\bibfnamefont{M.}~\bibnamefont{Sch\"utt}},
  \bibinfo{author}{\bibfnamefont{P.~M.} \bibnamefont{Ostrovsky}},
  \bibinfo{author}{\bibfnamefont{I.~V.} \bibnamefont{Gornyi}},
  \bibinfo{author}{\bibfnamefont{A.~D.} \bibnamefont{Mirlin}},
  \bibinfo{author}{\bibfnamefont{M.~I.} \bibnamefont{Katsnelson}},
  \bibinfo{author}{\bibfnamefont{K.~S.} \bibnamefont{Novoselov}},
  \bibnamefont{et~al.}, \bibinfo{journal}{Phys. Rev. Lett.}
  \textbf{\bibinfo{volume}{111}}, \bibinfo{pages}{166601}
  (\bibinfo{year}{2013}).

\bibitem[{\citenamefont{Alekseev et~al.}(2015)\citenamefont{Alekseev, Dmitriev,
  Gornyi, Kachorovskii, Narozhny, Sch\"utt, and Titov}}]{mr1}
\bibinfo{author}{\bibfnamefont{P.~S.} \bibnamefont{Alekseev}},
  \bibinfo{author}{\bibfnamefont{A.~P.} \bibnamefont{Dmitriev}},
  \bibinfo{author}{\bibfnamefont{I.~V.} \bibnamefont{Gornyi}},
  \bibinfo{author}{\bibfnamefont{V.~Y.} \bibnamefont{Kachorovskii}},
  \bibinfo{author}{\bibfnamefont{B.~N.} \bibnamefont{Narozhny}},
  \bibinfo{author}{\bibfnamefont{M.}~\bibnamefont{Sch\"utt}}, \bibnamefont{and}
  \bibinfo{author}{\bibfnamefont{M.}~\bibnamefont{Titov}},
  \bibinfo{journal}{Phys. Rev. Lett.} \textbf{\bibinfo{volume}{114}},
  \bibinfo{pages}{156601} (\bibinfo{year}{2015}).

\bibitem[{\citenamefont{Vasileva et~al.}(2016)\citenamefont{Vasileva, Smirnov,
  Ivanov, Vasilyev, Alekseev, Dmitriev, Gornyi, Kachorovskii, Titov, Narozhny
  et~al.}}]{mrexp}
\bibinfo{author}{\bibfnamefont{G.~Y.} \bibnamefont{Vasileva}},
  \bibinfo{author}{\bibfnamefont{D.}~\bibnamefont{Smirnov}},
  \bibinfo{author}{\bibfnamefont{Y.~L.} \bibnamefont{Ivanov}},
  \bibinfo{author}{\bibfnamefont{Y.~B.} \bibnamefont{Vasilyev}},
  \bibinfo{author}{\bibfnamefont{P.~S.} \bibnamefont{Alekseev}},
  \bibinfo{author}{\bibfnamefont{A.~P.} \bibnamefont{Dmitriev}},
  \bibinfo{author}{\bibfnamefont{I.~V.} \bibnamefont{Gornyi}},
  \bibinfo{author}{\bibfnamefont{V.~Y.} \bibnamefont{Kachorovskii}},
  \bibinfo{author}{\bibfnamefont{M.}~\bibnamefont{Titov}},
  \bibinfo{author}{\bibfnamefont{B.~N.} \bibnamefont{Narozhny}},
  \bibnamefont{et~al.}, \bibinfo{journal}{Phys. Rev. B}
  \textbf{\bibinfo{volume}{93}}, \bibinfo{pages}{195430}
  (\bibinfo{year}{2016}).

\bibitem[{\citenamefont{Ella et~al.}(2019)\citenamefont{Ella, Rozen, Birkbeck,
  Ben-Shalom, Perello, Zultak, Taniguchi, Watanabe, Geim, Ilani et~al.}}]{sulp}
\bibinfo{author}{\bibfnamefont{L.}~\bibnamefont{Ella}},
  \bibinfo{author}{\bibfnamefont{A.}~\bibnamefont{Rozen}},
  \bibinfo{author}{\bibfnamefont{J.}~\bibnamefont{Birkbeck}},
  \bibinfo{author}{\bibfnamefont{M.}~\bibnamefont{Ben-Shalom}},
  \bibinfo{author}{\bibfnamefont{D.}~\bibnamefont{Perello}},
  \bibinfo{author}{\bibfnamefont{J.}~\bibnamefont{Zultak}},
  \bibinfo{author}{\bibfnamefont{T.}~\bibnamefont{Taniguchi}},
  \bibinfo{author}{\bibfnamefont{K.}~\bibnamefont{Watanabe}},
  \bibinfo{author}{\bibfnamefont{A.~K.} \bibnamefont{Geim}},
  \bibinfo{author}{\bibfnamefont{S.}~\bibnamefont{Ilani}},
  \bibnamefont{et~al.}, \bibinfo{journal}{Nat. Nanotechnol.}
  \textbf{\bibinfo{volume}{14}}, \bibinfo{pages}{480} (\bibinfo{year}{2019}).

\bibitem[{\citenamefont{Ku et~al.}(2019)\citenamefont{Ku, Zhou, Li, Shin, Shi,
  Burch, Zhang, Casola, Taniguchi, Watanabe et~al.}}]{imh}
\bibinfo{author}{\bibfnamefont{M.~J.~H.} \bibnamefont{Ku}},
  \bibinfo{author}{\bibfnamefont{T.~X.} \bibnamefont{Zhou}},
  \bibinfo{author}{\bibfnamefont{Q.}~\bibnamefont{Li}},
  \bibinfo{author}{\bibfnamefont{Y.~J.} \bibnamefont{Shin}},
  \bibinfo{author}{\bibfnamefont{J.~K.} \bibnamefont{Shi}},
  \bibinfo{author}{\bibfnamefont{C.}~\bibnamefont{Burch}},
  \bibinfo{author}{\bibfnamefont{H.}~\bibnamefont{Zhang}},
  \bibinfo{author}{\bibfnamefont{F.}~\bibnamefont{Casola}},
  \bibinfo{author}{\bibfnamefont{T.}~\bibnamefont{Taniguchi}},
  \bibinfo{author}{\bibfnamefont{K.}~\bibnamefont{Watanabe}},
  \bibnamefont{et~al.} (\bibinfo{year}{2019}),
  \bibinfo{note}{arXiv:1905.10791}.

\bibitem[{\citenamefont{Sulpizio et~al.}(2019)\citenamefont{Sulpizio, Ella,
  Rozen, Birkbeck, Perello, Dutta, Ben-Shalom, Taniguchi, Watanabe, Holder
  et~al.}}]{imm}
\bibinfo{author}{\bibfnamefont{J.~A.} \bibnamefont{Sulpizio}},
  \bibinfo{author}{\bibfnamefont{L.}~\bibnamefont{Ella}},
  \bibinfo{author}{\bibfnamefont{A.}~\bibnamefont{Rozen}},
  \bibinfo{author}{\bibfnamefont{J.}~\bibnamefont{Birkbeck}},
  \bibinfo{author}{\bibfnamefont{D.~J.} \bibnamefont{Perello}},
  \bibinfo{author}{\bibfnamefont{D.}~\bibnamefont{Dutta}},
  \bibinfo{author}{\bibfnamefont{M.}~\bibnamefont{Ben-Shalom}},
  \bibinfo{author}{\bibfnamefont{T.}~\bibnamefont{Taniguchi}},
  \bibinfo{author}{\bibfnamefont{K.}~\bibnamefont{Watanabe}},
  \bibinfo{author}{\bibfnamefont{T.}~\bibnamefont{Holder}},
  \bibnamefont{et~al.}, \bibinfo{journal}{Nature}
  \textbf{\bibinfo{volume}{576}}, \bibinfo{pages}{75} (\bibinfo{year}{2019}).

\bibitem[{\citenamefont{Lifshitz and Pitaevskii}(1981)}]{dau10}
\bibinfo{author}{\bibfnamefont{E.~M.} \bibnamefont{Lifshitz}} \bibnamefont{and}
  \bibinfo{author}{\bibfnamefont{L.~P.} \bibnamefont{Pitaevskii}},
  \emph{\bibinfo{title}{Physical Kinetics}} (\bibinfo{publisher}{Pergamon
  Press, London}, \bibinfo{year}{1981}).

\bibitem[{\citenamefont{Narozhny et~al.}(2001)\citenamefont{Narozhny, Aleiner,
  and Stern}}]{df2}
\bibinfo{author}{\bibfnamefont{B.~N.} \bibnamefont{Narozhny}},
  \bibinfo{author}{\bibfnamefont{I.~L.} \bibnamefont{Aleiner}},
  \bibnamefont{and} \bibinfo{author}{\bibfnamefont{A.}~\bibnamefont{Stern}},
  \bibinfo{journal}{Phys. Rev. Lett.} \textbf{\bibinfo{volume}{86}},
  \bibinfo{pages}{3610} (\bibinfo{year}{2001}).

\bibitem[{\citenamefont{Ziman}(1965)}]{ziman}
\bibinfo{author}{\bibfnamefont{J.~M.} \bibnamefont{Ziman}},
  \emph{\bibinfo{title}{Principles of the Theory of Solids}}
  (\bibinfo{publisher}{Cambridge University Press, Cambridge},
  \bibinfo{year}{1965}).

\bibitem[{\citenamefont{Giuliani and Vignale}(2005)}]{Giuliani}
\bibinfo{author}{\bibfnamefont{G.}~\bibnamefont{Giuliani}} \bibnamefont{and}
  \bibinfo{author}{\bibfnamefont{G.}~\bibnamefont{Vignale}},
  \emph{\bibinfo{title}{Quantum Theory of the Electron Liquid}}
  (\bibinfo{publisher}{Cambridge University Press}, \bibinfo{year}{2005}).

\bibitem[{\citenamefont{Katsnelson}(2012)}]{kats}
\bibinfo{author}{\bibfnamefont{M.~I.} \bibnamefont{Katsnelson}},
  \emph{\bibinfo{title}{Graphene}} (\bibinfo{publisher}{Cambridge University
  Press}, \bibinfo{year}{2012}).

\bibitem[{\citenamefont{Narozhny}(2019)}]{me1}
\bibinfo{author}{\bibfnamefont{B.~N.} \bibnamefont{Narozhny}},
  \bibinfo{journal}{Annals of Physics} \textbf{\bibinfo{volume}{411}},
  \bibinfo{pages}{167979} (\bibinfo{year}{2019}).

\bibitem[{\citenamefont{Aleiner and Shklovskii}(1994)}]{ash}
\bibinfo{author}{\bibfnamefont{I.~L.} \bibnamefont{Aleiner}} \bibnamefont{and}
  \bibinfo{author}{\bibfnamefont{B.~I.} \bibnamefont{Shklovskii}},
  \bibinfo{journal}{Phys. Rev. B} \textbf{\bibinfo{volume}{49}},
  \bibinfo{pages}{13721} (\bibinfo{year}{1994}).

\bibitem[{\citenamefont{Shik}(1993)}]{shik}
\bibinfo{author}{\bibfnamefont{A.}~\bibnamefont{Shik}}, \bibinfo{journal}{J.
  Phys. Condens. Matter} \textbf{\bibinfo{volume}{5}}, \bibinfo{pages}{8963}
  (\bibinfo{year}{1993}).

\bibitem[{\citenamefont{Foster and Aleiner}(2009)}]{alf}
\bibinfo{author}{\bibfnamefont{M.~S.} \bibnamefont{Foster}} \bibnamefont{and}
  \bibinfo{author}{\bibfnamefont{I.~L.} \bibnamefont{Aleiner}},
  \bibinfo{journal}{Phys. Rev. B} \textbf{\bibinfo{volume}{79}},
  \bibinfo{pages}{085415} (\bibinfo{year}{2009}).

\bibitem[{\citenamefont{Xie and Levchenko}(2019)}]{lev19}
\bibinfo{author}{\bibfnamefont{H.-Y.} \bibnamefont{Xie}} \bibnamefont{and}
  \bibinfo{author}{\bibfnamefont{A.}~\bibnamefont{Levchenko}},
  \bibinfo{journal}{Phys. Rev. B} \textbf{\bibinfo{volume}{99}},
  \bibinfo{pages}{045434} (\bibinfo{year}{2019}).

\bibitem[{\citenamefont{Kiselev and Schmalian}(2019)}]{ks19}
\bibinfo{author}{\bibfnamefont{E.~I.} \bibnamefont{Kiselev}} \bibnamefont{and}
  \bibinfo{author}{\bibfnamefont{J.}~\bibnamefont{Schmalian}},
  \bibinfo{journal}{Phys. Rev. B} \textbf{\bibinfo{volume}{99}},
  \bibinfo{pages}{035430} (\bibinfo{year}{2019}).

\bibitem[{\citenamefont{Gallagher et~al.}(2019)\citenamefont{Gallagher, Yang,
  Lyu, Tian, Kou, Zhang, Watanabe, Taniguchi, and Wang}}]{gal}
\bibinfo{author}{\bibfnamefont{P.}~\bibnamefont{Gallagher}},
  \bibinfo{author}{\bibfnamefont{C.-S.} \bibnamefont{Yang}},
  \bibinfo{author}{\bibfnamefont{T.}~\bibnamefont{Lyu}},
  \bibinfo{author}{\bibfnamefont{F.}~\bibnamefont{Tian}},
  \bibinfo{author}{\bibfnamefont{R.}~\bibnamefont{Kou}},
  \bibinfo{author}{\bibfnamefont{H.}~\bibnamefont{Zhang}},
  \bibinfo{author}{\bibfnamefont{K.}~\bibnamefont{Watanabe}},
  \bibinfo{author}{\bibfnamefont{T.}~\bibnamefont{Taniguchi}},
  \bibnamefont{and} \bibinfo{author}{\bibfnamefont{F.}~\bibnamefont{Wang}},
  \bibinfo{journal}{Science} \textbf{\bibinfo{volume}{364}},
  \bibinfo{pages}{158} (\bibinfo{year}{2019}).

\bibitem[{\citenamefont{Narozhny et~al.}(2012)\citenamefont{Narozhny, Titov,
  Gornyi, and Ostrovsky}}]{drag}
\bibinfo{author}{\bibfnamefont{B.~N.} \bibnamefont{Narozhny}},
  \bibinfo{author}{\bibfnamefont{M.}~\bibnamefont{Titov}},
  \bibinfo{author}{\bibfnamefont{I.~V.} \bibnamefont{Gornyi}},
  \bibnamefont{and} \bibinfo{author}{\bibfnamefont{P.~M.}
  \bibnamefont{Ostrovsky}}, \bibinfo{journal}{Phys. Rev. B}
  \textbf{\bibinfo{volume}{85}}, \bibinfo{pages}{195421}
  (\bibinfo{year}{2012}).

\bibitem[{\citenamefont{Ando}(2006)}]{ando}
\bibinfo{author}{\bibfnamefont{T.}~\bibnamefont{Ando}}, \bibinfo{journal}{J.
  Phys. Soc. Jpn.} \textbf{\bibinfo{volume}{75}}, \bibinfo{pages}{074716}
  (\bibinfo{year}{2006}).

\bibitem[{\citenamefont{Shon and Ando}(1998)}]{ando2}
\bibinfo{author}{\bibfnamefont{N.}~\bibnamefont{Shon}} \bibnamefont{and}
  \bibinfo{author}{\bibfnamefont{T.}~\bibnamefont{Ando}}, \bibinfo{journal}{J.
  Phys. Soc. Jpn.} \textbf{\bibinfo{volume}{67}}, \bibinfo{pages}{2421}
  (\bibinfo{year}{1998}).

\bibitem[{\citenamefont{Nomura and MacDonald}(2006)}]{nom}
\bibinfo{author}{\bibfnamefont{K.}~\bibnamefont{Nomura}} \bibnamefont{and}
  \bibinfo{author}{\bibfnamefont{A.~H.} \bibnamefont{MacDonald}},
  \bibinfo{journal}{Phys. Rev. Lett.} \textbf{\bibinfo{volume}{96}},
  \bibinfo{pages}{256602} (\bibinfo{year}{2006}).

\bibitem[{\citenamefont{Cheianov and Fal'ko}(2006)}]{fal}
\bibinfo{author}{\bibfnamefont{V.~V.} \bibnamefont{Cheianov}} \bibnamefont{and}
  \bibinfo{author}{\bibfnamefont{V.~I.} \bibnamefont{Fal'ko}},
  \bibinfo{journal}{Phys. Rev. Lett.} \textbf{\bibinfo{volume}{97}},
  \bibinfo{pages}{226801} (\bibinfo{year}{2006}).

\bibitem[{\citenamefont{Aleiner and Efetov}(2006)}]{alef}
\bibinfo{author}{\bibfnamefont{I.~L.} \bibnamefont{Aleiner}} \bibnamefont{and}
  \bibinfo{author}{\bibfnamefont{K.~B.} \bibnamefont{Efetov}},
  \bibinfo{journal}{Phys. Rev. Lett.} \textbf{\bibinfo{volume}{97}},
  \bibinfo{pages}{236801} (\bibinfo{year}{2006}).

\bibitem[{\citenamefont{Ostrovsky et~al.}(2006)\citenamefont{Ostrovsky, Gornyi,
  and Mirlin}}]{ogm}
\bibinfo{author}{\bibfnamefont{P.~M.} \bibnamefont{Ostrovsky}},
  \bibinfo{author}{\bibfnamefont{I.~V.} \bibnamefont{Gornyi}},
  \bibnamefont{and} \bibinfo{author}{\bibfnamefont{A.~D.}
  \bibnamefont{Mirlin}}, \bibinfo{journal}{Phys. Rev. B}
  \textbf{\bibinfo{volume}{74}}, \bibinfo{pages}{235443}
  (\bibinfo{year}{2006}).

\bibitem[{\citenamefont{Chiappini et~al.}(2016)\citenamefont{Chiappini,
  Wiedmann, Titov, Geim, Gorbachev, Khestanova, Mishchenko, Novoselov, Maan,
  and Zeitler}}]{chiap}
\bibinfo{author}{\bibfnamefont{F.}~\bibnamefont{Chiappini}},
  \bibinfo{author}{\bibfnamefont{S.}~\bibnamefont{Wiedmann}},
  \bibinfo{author}{\bibfnamefont{M.}~\bibnamefont{Titov}},
  \bibinfo{author}{\bibfnamefont{A.~K.} \bibnamefont{Geim}},
  \bibinfo{author}{\bibfnamefont{R.~V.} \bibnamefont{Gorbachev}},
  \bibinfo{author}{\bibfnamefont{E.}~\bibnamefont{Khestanova}},
  \bibinfo{author}{\bibfnamefont{A.}~\bibnamefont{Mishchenko}},
  \bibinfo{author}{\bibfnamefont{K.~S.} \bibnamefont{Novoselov}},
  \bibinfo{author}{\bibfnamefont{J.~C.} \bibnamefont{Maan}}, \bibnamefont{and}
  \bibinfo{author}{\bibfnamefont{U.}~\bibnamefont{Zeitler}},
  \bibinfo{journal}{Phys. Rev. B} \textbf{\bibinfo{volume}{94}},
  \bibinfo{pages}{085302} (\bibinfo{year}{2016}).

\end{thebibliography}

\end{document}